\documentclass[twocolumn]{aastex62}

\received{ }
\revised{ }
\accepted{ }
\submitjournal{ApJ}

\shorttitle{A wildly flickering jet in MAXI J1535--571}
\shortauthors{Baglio, M. C. et al.}

\begin{document}

\title{A wildly flickering jet in the black hole X-ray binary MAXI J1535--571}

\correspondingauthor{Maria Cristina Baglio}
\email{mcb19@nyu.edu}

\author[0000-0003-1285-4057]{Maria Cristina Baglio}
\affiliation{New York University Abu Dhabi, PO Box 129188, Abu Dhabi, UAE}
\affiliation{INAF, Osservatorio Astronomico di Brera, Via E. Bianchi 46, I-23807 Merate (LC), Italy}

\author{David M. Russell}
\affiliation{New York University Abu Dhabi, PO Box 129188, Abu Dhabi, UAE}

\author{Piergiorgio Casella}
\affiliation{INAF, Osservatorio Astronomico di Roma, Via Frascati 33, I-00040, Monteporzio Catone (RM), Italy}

\author{Hind Al Noori}
\affiliation{New York University Abu Dhabi, PO Box 129188, Abu Dhabi, UAE}

\author{Aisha Al Yazeedi}
\affiliation{New York University Abu Dhabi, PO Box 129188, Abu Dhabi, UAE}

\author{Tomaso Belloni}
\affiliation{INAF, Osservatorio Astronomico di Brera, Via E. Bianchi 46, I-23807 Merate (LC), Italy}

\author{David A. H. Buckley}
\affiliation{South African Astronomical Observatory, P.O. Box 9, Observatory 7935, Cape Town, South Africa}

\author{Marion Cadolle Bel}
\affiliation{SIXT Leasing SE, Zugspitzstr. 1, 82049 Pullach (Munich), Germany}

\author{Chiara Ceccobello}
\affiliation{Department of Space, Earth and Environment, Chalmers University of Technology, Onsala Space Observatory, 439 92 Onsala, Sweden}

\author{Stephane Corbel}
\affiliation{Laboratoire AIM (CEA/IRFU - CNRS/INSU - Universit\'{e} Paris Diderot), CEA DSM/IRFU/SAp, F-91191 Gif-sur-Yvette, France}
\affiliation{Station de Radioastronomie de Nan\c{c}ay, Observatoire de Paris, PSL Research University, CNRS, Univ. Orl\'{e}ans, F-18330 Nan\c{c}ay, France}

\author{Francesco Coti Zelati}
\affiliation{Institute of Space Sciences (ICE, CSIC), Campus UAB, Carrer de Can Magrans s/n, E-08193 Barcelona, Spain}
\affiliation{Institut dÕEstudis Espacials de Catalunya (IEEC), E-08034 Barcelona, Spain}
\affiliation{INAF, Osservatorio Astronomico di Brera, Via E. Bianchi 46, I-23807 Merate (LC), Italy}

\author{Maria D\'{i}az Trigo}
\affiliation{ESO, Karl-Schwarzschild-Strasse 2, D-85748 Garching bei M\"unchen, Germany}

\author {Rob P. Fender} 
\affiliation{Astrophysics, Department of Physics, University of Oxford, Keble Road, Oxford OX1 3RH, UK}

\author{Elena Gallo}
\affiliation{Department of Astronomy, University of Michigan, 1085 S University, Ann Arbor, MI 48109, USA}

\author{Poshak Gandhi}
\affiliation{University of Southampton, Department of Physics \& Astronomy, Southampton, SO17 1BJ, UK}

\author{Jeroen Homan}
\affiliation{Eureka Scientific, Inc., 2452 Delmer Street, Oakland, CA 94602, USA}
\affiliation{SRON, Netherlands Institute for Space Research, Sorbonnelaan 2, 3584 CA Utrecht, The Netherlands}

\author {Karri I. I. Koljonen}
\affiliation{Finnish Centre for Astronomy with ESO (FINCA), University of Turku, V\"ais\"al\"antie 20, 21500 Piikki\"o, Finland}
\affiliation{Aalto University Mets\"ahovi Radio Observatory, PO Box 13000, FI-00076 Aalto, Finland}\

\author{Fraser Lewis}
\affiliation{Faulkes Telescope Project, School of Physics and Astronomy, Cardiff University, The Parade, Cardiff CF24 3AA, UK}
\affiliation{Astrophysics Research Institute, Liverpool John Moores University, 146 Brownlow Hill, Liverpool L3 5RF, UK}

\author {Thomas J. Maccarone}
\affiliation{Department of Physics and Astronomy, Texas Tech University, Box 41051, Lubbock, TX 79409-1051, USA}

\author{Julien Malzac}
\affiliation{IRAP, UniversitŽ de Toulouse, CNRS, UPS, CNES, Toulouse, France}

\author{Sera Markoff}
\affiliation{Anton Pannekoek Institute for Astronomy, University of Amsterdam, 1098 XH Amsterdam, The Netherlands}

\author{James C.A. Miller-Jones}
\affiliation{International Centre for Radio Astronomy Research, Curtin University, GPO Box U1987, Perth, Western Australia 6845, Australia}

\author{Kieran O'Brien}
\affiliation{Department of Physics, Durham University, South Road, Durham, DH1 3LE, UK}

\author{Thomas D. Russell}
\affiliation{Anton Pannekoek Institute for Astronomy, University of Amsterdam, 1098 XH Amsterdam, The Netherlands}

\author{Payaswini Saikia}
\affiliation{New York University Abu Dhabi, PO Box 129188, Abu Dhabi, UAE}

\author{Tariq Shahbaz}
\affiliation{Instituto de Astrof\'\i{}sica de Canarias (IAC), E-38200 La Laguna, Tenerife, Spain}
\affiliation{Departamento de  Astrof\'\i{}sica, Universidad de La Laguna (ULL), E-38206 La Laguna, Tenerife, Spain}

\author{Greg R. Sivakoff}
\affiliation{Department of Physics, University of Alberta, CCIS 4-181, Edmonton, AB T6G 2E1, Canada}

\author{Roberto Soria}
\affiliation{College of Astronomy and Space Sciences, University of Chinese Academy of Sciences, Beijing 100049, China}
\affiliation{International Centre for Radio Astronomy Research, Curtin University, GPO Box U1987, Perth, Western Australia 6845, Australia}
\affiliation{Sydney Institute for Astronomy, School of Physics A28, The University of Sydney, Sydney, NSW 2006, Australia}

\author{Vincenzo Testa}
\affiliation{INAF, Osservatorio Astronomico di Roma, Via Frascati 33, I-00040, Monteporzio Catone (RM), Italy}

\author{Alexandra J. Tetarenko}
\affiliation{Department of Physics, University of Alberta, CCIS 4-181, Edmonton, AB T6G 2E1, Canada}
\affiliation{East Asian Observatory, 660 N. A'ohoku Place, University Park, Hilo, Hawaii 96720, USA}

\author {Mario E. van den Ancker}
\affiliation{ESO, Karl-Schwarzschild-Strasse 2, D-85748 Garching bei M\"unchen, Germany}

\author{Federico M. Vincentelli}
\affiliation{DiSAT, Universit\`{a} degli Studi dellÕInsubria, Via Valleggio 11, I-22100 Como, Italy}
\affiliation{INAF, Osservatorio Astronomico di Brera, Via E. Bianchi 46, I-23807 Merate (LC), Italy}
\affiliation{INAF, Osservatorio Astronomico di Roma, Via Frascati 33, I-00040, Monteporzio Catone (RM), Italy}

\begin{abstract}
We report on the results of optical, near-infrared (NIR) and mid-infrared observations of the  black hole X-ray binary candidate (BHB) MAXI J1535--571 during its 2017/2018 outburst.
During the first part of the outburst (MJD 58004-58012), the source shows an optical-NIR spectrum that is consistent with an optically thin synchrotron power-law from a jet. After MJD 58015, however, the source faded considerably, the drop in flux being much more evident at lower frequencies. Before the fading, we measure a de-reddened flux density of $\gtrsim$100 mJy in the mid-infrared, making MAXI J1535--571 one of the brightest mid-infrared BHBs known so far. A significant softening of the X-ray spectrum is evident contemporaneous with the infrared fade. We interpret it as due to the suppression of the jet emission, similar to the accretion-ejection coupling seen in other BHBs. However, MAXI J1535--571 did not transition smoothly to the soft state, instead showing X-ray hardness deviations, associated with infrared flaring. We also present the first mid-IR variability study of a BHB on minute timescales, with a fractional rms variability of the light curves of $\sim 15$--22 \%, which is similar to that expected from the internal shock jet model, and much higher than the optical fractional rms ($\lesssim 7$ \%).
These results represent an excellent case of multi-wavelength jet spectral-timing and demonstrate how rich, multi-wavelength time-resolved data of X-ray binaries over accretion state transitions can help refining models of the disk-jet connection and jet launching in these systems.
\end{abstract}



\section{Introduction} \label{intro}

Low-mass X-ray binaries (LMXBs) are systems that typically host an evolved/main-sequence star and a compact object, either a neutron star (NS) or a black hole (BH). The companion star of these systems transfers mass and angular momentum toward the compact object via Roche lobe overflow and the formation of an accretion disk. LMXBs can be transient, which means that they alternate between outbursts, typically lasting weeks to months, with X-ray luminosities reaching $ 10^{36}-10^{38} \, \rm erg/s $ and high accretion levels, and longer (years/decades) periods of quiescence, with a typical drop in the X-ray luminosity by 5-6 orders of magnitude.

A coupling between accretion and ejection of matter is thought to exist for LMXBs (\citealt{Merloni2003}; \citealt{Falcke2004}; \citealt{Plotkin2013}). These ejections are mainly seen in the form of collimated and compact jets coming from a region that is very near to the central compact object. 
The production of relativistic jets in black hole X-ray binaries (BHBs) is a matter of extensive study in the field of accreting systems. 
The magnetic field in the inner part of the accretion flow is thought to be responsible for particle acceleration and jet launching. In the case of jet emission, a flat radio spectrum is typically observed when the system is in its hard X-ray state (\citealt{Fender01}; \citealt{FenderBelloniGallo04}; \citealt{Corbel2004}). This flat or inverted ($\alpha \sim 0-0.5$, with $F_{\rm \nu}\propto \nu^{\rm \alpha}$) spectrum is one of the principal signatures of the presence of a collimated jet in BHBs. It typically extends at least to the millimeter regime, and is normally explained as due to the superposition of self-absorbed synchrotron emission components originating at different distances from the central BH. At higher frequencies, often starting from the mid-IR (MIR) or near-infrared (NIR) bands, optically thin synchrotron emission is observed, resulting in a power-law with index $-1\leq \alpha \leq -0.5$ \citep{Russell2013a}. The transition between the optically thick and thin parts of the spectrum happens at the so-called \textit{jet break} frequency, $\nu_{\rm b}$, that usually falls in the MIR to NIR \citep{Russell2013a}. In the soft X-ray state instead, jets appear to be quenched at all frequencies (see e.g. \citealt{Russell2011quenching}) in favor of disk winds \citep{Ponti2012}, probably due to the suppression of the magnetic field caused by the geometrically thin accretion disk, that resides close to the black hole \citep[see e.g.][]{Meier2001}. Alternatively, in line with the internal shock model for BHBs jets by \citet{Malzac2013}, during the soft state the jet might still be present, but dark due to the lack of variability in the disk \citep{Drappeau2017}. 
In some cases, the jet break frequency has been found to move from infrared to radio frequencies as the X-ray spectrum of the source softens, and then to come back again to the infrared band at the end of the outburst, when the spectrum gets harder (\citealt{Russell2013b}; \citealt{Russell2014}; \citealt{DiazTrigo2018}). This behavior suggests that the structure of the accretion flow determines the position of the break frequency of the jet.

The size of the jet emitting region is thought to scale inversely with the frequency \citep{Blandford79}; in this way, $\nu_{\rm b}$ would give information on the size of the base of the jet, i.e. where the particle acceleration starts (\citealt{Heinz2003}; \citealt{Chaty2011}; \citealt{Ceccobello2018}). Moreover, the detection of the jet break frequency is fundamental also to estimate the magnetic field strength and the total power of the jet and its radiative luminosity, and therefore to shed light on the jet formation process and the inflow/outflow connection in X-ray binaries \citep{Russell2014}. The break has been detected in the case of the BH candidates ~GX~339--4  (\citealt{Corbel02}; \citealt{Gandhi11}; \citealt{Corbel2013}), XTE~J1118+480 \citep{Hynes2006}, XTE~J1550--564 \citep{Chaty2011}, V404~Cyg \citep{Russell2013a,Tetarenko2018}, MAXI~J1836--194 \citep{Russell2013b,Russell2014}, Cyg~X--1 \citep{Rahoui2011}, and in quiescence, Swift~J1357.2--0933 and A0620--00 (\citealt{Plotkin2016}; \citealt{Russell2018}; \citealt{Dincer2018}) (and in the neutron star systems 4U~0614+091, 4U 1728--34 and Aql~X--1; \citealt{Migliari10}, \citealt{DiazTrigo2017}, \citealt{DiazTrigo2018}).

\section{MAXI J1535--571}\label{sec_maxi}
The BHB MAXI J1535--571 (hereafter J1535) was first detected by the \textit{MAXI}/GSC nova alert system as a bright, hard X-ray transient on September 2nd, 2017 \citep{ATEL10699}. Independently, the {\it Swift}/Burst Alert Telescope ({\it Swift}/BAT) instrument triggered on an outburst from a possible Gamma Ray Burst (\citealt{ATEL10700}), coming from the same position, which led to the conclusion that they were observing the same event. The spectrum of the source was well fit by an absorbed power-law with a hydrogen column density $N_{\rm H}=(3.6\pm0.2)\,\times 10^{22}\,\rm cm^{-2}$ and a photon index of 1.53$\pm$0.07 \citep{ATEL10700}. 

The optical counterpart was first detected on September 3rd using the 0.61m B\&C Telescope operated by the University of Canterbury at Mt. John Observatory, located in New Zealand, which revealed a $i'=21.8\pm0.2$ mag source inside the {\it Swift}/X-ray telescope ({\it Swift}/XRT) error circle \citep{ATEL10702}. The target was not detected at higher frequencies (i.e. $g'$ and $r'$ bands), not surprisingly due to the high extinction at that location in the Galactic Plane.

In the 2--20 keV range, the X-ray flux increased linearly following the first detection \citep{ATEL10708}. 
From further \textit{MAXI}/GSC observations reported in \citet{ATEL10708}, the unabsorbed 1--60 keV flux ($3.1\times10^{-8}\, \rm erg/s/cm^{2}$) corresponded to a luminosity of $2.4\times10^{38}$ erg/s assuming a 8 kpc distance, which exceeds the Eddington luminosity of a canonical 1.4$M_{\odot}$ neutron star by a factor of $\sim 1.4$. Moreover, rapid X-ray variability without clear periodicity is observed in the GSC light curve. Due to these observed properties, the authors suggested that the source could be a hard-state low-mass X-ray binary hosting a black hole.

Observations with the \textit{Australia Telescope Compact Array} (\textit{ATCA}) were performed \citep{ATEL10711} on September 5th, at 5.5 and 9 GHz. The authors detected a radio source at a position consistent with the X-ray one, with a radio spectral index of $\alpha=0.09\pm0.03$, which is consistent with emission from a compact radio jet. They also estimated the distance of the target to be about 6.5 kpc, assuming the source to be as close to the Galactic centre as possible along its line-of-sight. The observed radio emission is well above the expected radio luminosity of a neutron star low-mass X-ray binary at the observed X-ray luminosity, whereas it is comparable to those of typical BHBs, indicating that J1535 is indeed a strong BHB candidate. Successively, the spectral and timing analysis of archival {\it Swift} (BAT and XRT) and \textit{MAXI} (GSC) data reported in \citet{Shang2018} led to an estimate of the probable BH mass of $8.78^{+1.22}_{-1.05}$ $ M_{\odot}$.

The NIR counterpart has been detected with the SMARTS 1.3m telescope at CTIO as a J=14.88, H=13.11 mag source, on September 5th \citep{ATEL10716}.

The \textit{MAXI}/GSC and {\it Swift} monitoring of the outburst revealed a linearly increasing trend of the X-ray luminosity with time, the 2--20 keV flux reaching $5.3\times10^{-8}\, \rm erg\, \rm cm^{-2}\,\rm s^{-1}$ on September 10th. After this date, while the \textit{MAXI} 2--4 keV and the XRT count rate continued increasing, the \textit{MAXI} 10--20 keV and the BAT count rates started to decline, while the X-ray photon index became softer, indicating a possible hard-soft transition of the source (\citealt{ATEL10729}, \citealt{ATEL10731}, \citealt{ATEL10733}). 

On September 11th, millimeter observations with the \textit{Atacama Large Millimeter/Sub-millimeter array} (\textit{ALMA}) were performed. \textit{ALMA} observations taken at 97, 140, and 230 GHz marked J1535 as one of the brightest X-ray binaries ever detected at sub-mm wavelengths \citep{ATEL10745}. \textit{ATCA} observations performed on September 12th confirmed the brightening of the source in the radio band, both at 17 and 19 GHz \citep{ATEL10745}. The authors linked these radio detections to synchrotron emission arising from a compact jet, with a spectrum consistent with a single power-law extending from the radio up to $\sim$140 GHz, with the break residing above that.

A softening of the X-ray spectrum on September 19th (MJD 58015) is first reported by \citet{ATEL10761}, suggesting that the source entered the so-called soft- intermediate state \citep{Tao2018}. 
A further brightening in the radio was then detected with \textit{ATCA} on October 25th \citep{ATEL10899}, with a radio spectrum consistent with a single power-law (spectral index of $\sim 0.09\pm 0.01$), again indicative of synchrotron emission from a compact jet. This suggested that the source may have been transitioning back towards the hard state.

\citet{ATEL11020} reported on a steady spectral softening of the source after October 25th, reaching a hardness ratio of the \textit{MAXI}/GSC fluxes (6--20 keV/2--6 keV) of $\sim$0.03 in the last ten days until November 27th. This is the lowest hardness ratio reached by J1535 during the outburst, and is significantly lower than the one measured soon before the temporary hardening that occurred at the end of October ($\sim 0.1$), thus indicating that the source finally reached the soft state, where it remained for several months (see Section \ref{HID_section} and the work by \citealt{Tao2018} for a detailed description of the evolution of the X-ray outburst of J1535).

Here, we present optical, NIR and MIR observations of J1535 during the outburst rise and period of spectral transitions. We use the evolution of the emission properties to probe the disk--jet connection in this source. We also present the first MIR rapid (minute timescale) variations reported in a transient BHB.

\section{Observations}

\subsection{Optical and NIR monitoring}
J1535 was extensively monitored during its 2017 outburst using the 2-m Faulkes Telescope South (FTS) located at Siding Spring (Australia), the Las Cumbres Observatory (LCO) robotic network of 1-m telescopes at Cerro Tololo Inter-American Observatory in Chile (LSC), the South African Astronomical Observatory at Sutherland in South Africa (CPT) and Siding Spring (COJ), and the 1-m ESO Rapid Eye Mount telescope (REM) in La Silla (Chile).

\subsubsection{Faulkes/LCO observations}

We observed J1535 with FTS and the 1-m LCO network on 20 nights in September and October 2017, since the first optical detection until the source became no longer visible from the ground. These observations are part of an ongoing monitoring campaign of $\sim 40$ LMXBs \citep{Lewis08}. Imaging data were taken in the SDSS $g'r'i'$ and Pan-STARRS $y$-band filters (4770--10040$\AA$). We detected the source (with errors of $\leq 0.4$ mag) in 0, 1, 12 and 11 images in $g'$, $r'$, $i'$ and $y$-bands, respectively. As for the REM observations, the non-detections at shorter wavelengths are very likely due to the high foreground extinction, and led us to schedule only $i'$ and $y$-band observations from September 12th. A detailed observation log can be found in Tab. \ref{LCO_log}.

\begin{figure}
\centering
\includegraphics[scale=0.35]{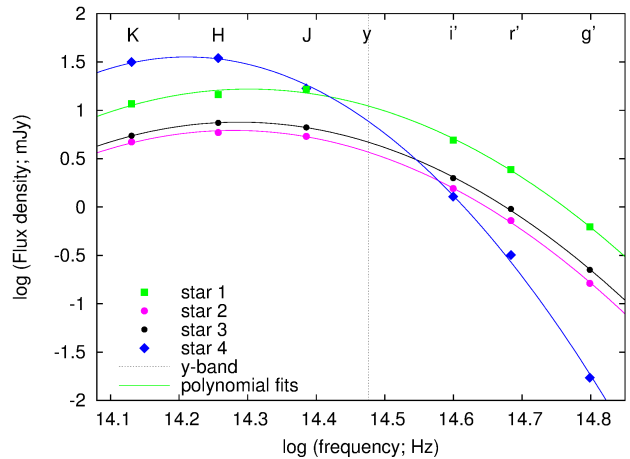}
\caption{Polynomial fits to the optical--infrared SEDs of four field stars, used to calibrate the $y$-band LCO data. The vertical dotted line corresponds to the central frequency of the $y$-band filter.}
\label{calib_y}
\end{figure}

The Faulkes/LCO data were reduced (de-biased and flat-fielded) using the LCO automatic pipeline. Photometry was performed using \textsc{PHOT} in \textsc{IRAF}.\footnote{IRAF is distributed by the National Optical Astronomy Observatory, which is operated by the Association of Universities for Research in Astronomy, Inc., under cooperative agreement with the National Science Foundation.} For photometric calibration of the $r',i'$-bands we used four close by stars listed in the VST Photometric H$\alpha$ Survey of the Southern Galactic Plane and Bulge (VPHAS+) Data Release 2 \citep{Drew14,Drew16}. The $r$,$i$ Vega catalogue magnitudes of these stars were converted to $r'$, $i'$ AB magnitudes, which is standard for SDSS filters. We found no catalogues with $y$-band magnitudes of any field stars (Pan-STARRS does not cover this field). However, we performed log-log polynomial fits to the $g',r',i',J,H,K$ spectral energy distributions (SEDs) of these four field stars using the VPHAS+ and 2MASS catalogues, and interpolated the fluxes of each star to the $y$-band frequency in order to estimate the $y$-band magnitudes of the four stars. The polynomial curves provided good approximations to the reddened, black body SEDs of the stars (Fig. \ref{calib_y}). In all filters, the errors are dominated by the S/N of the target (not the systematic calibration uncertainty).

\begin{table}[htb]
\caption{Detailed log of the Faulkes/LCO observations performed between Sep. 4, 2017 and Oct. 13, 2017. FTS is the 2-m Faulkes Telescope South in Australia, 1m-A is a 1-m node in Australia, 1m-C is a 1-m node in Chile and 1m-S is a 1-m node in South Africa.}            
\label{LCO_log}      
\centering                       
\begin{tabular}{c |c| c| c| c}       
\hline\hline                
Epoch & MJD & Tele- & Filters & Exposures \\   
 (UT) &     & scope &        & (s) \\
\hline                     
2017-09-04 & 58000.37035 & FTS  & $r'$   & $3 \times 300$ \\
2017-09-06 & 58002.42514 & FTS  & $i'$   & 200 \\
2017-09-06 & 58002.78674 & 1m-S & $g'i'$ & 300,200 \\
2017-09-07 & 58003.42129 & 1m-A & $g'i'$ & 300,200 \\
2017-09-07 & 58003.78054 & 1m-S & $g'i'$ & 300,200 \\
2017-09-07 & 58003.78771 & 1m-S & $r'y$  & 200,200 \\
2017-09-08 & 58004.38974 & FTS  & $r'$   & 300 \\
2017-09-08 & 58004.98666 & 1m-C & $y$    & 200 \\
2017-09-09 & 58005.44592 & FTS  & $i'$   & 200 \\
2017-09-09 & 58005.76490 & 1m-S & $y$    & 200 \\
2017-09-10 & 58006.43538 & FTS  & $i'$   & 200 \\
2017-09-10 & 58006.77602 & 1m-S & $g'i'y$& 300,200,200 \\
2017-09-11 & 58007.42086 & FTS  & $r'i'$ & $2 \times 300$,200 \\
2017-09-12 & 58008.42079 & FTS  & $i'$   & 200 \\
2017-09-13 & 58009.36657 & FTS  & $i'$   & 200 \\
2017-09-13 & 58009.43914 & 1m-A & $y$    & 200 \\
2017-09-14 & 58010.79202 & 1m-S & $y$    & 200 \\
2017-09-21 & 58017.40310 & 1m-A & $i'y$  & 300,300 \\
2017-09-21 & 58017.43811 & FTS  & $i'y$  & 300,300 \\
2017-09-22 & 58018.40711 & 1m-A & $i'y$  & 300,300 \\
2017-09-23 & 58019.01422 & 1m-C & $i'y$  & 300,300 \\
2017-09-25 & 58021.42373 & FTS  & $i'y$  & 300,300 \\
2017-09-25 & 58021.98683 & 1m-C & $i'y$  & 300,300 \\
2017-10-02 & 58028.99327 & 1m-C & $i'y$  & 200,200 \\
2017-10-03 & 58029.37504 & 1m-A & $i'y$  & 200,200 \\
2017-10-04 & 58030.98582 & 1m-C & $i'y$  & 200,200 \\
2017-10-06 & 58032.37651 & 1m-A & $i'y$  & 200,200 \\
2017-10-07 & 58033.41709 & FTS  & $i'y$  & 300,300 \\
2017-10-13 & 58039.39960 & FTS  & $i'y$  & 300,300 \\
\hline                             
\end{tabular}
\end{table}

\subsubsection{REM observations}
J1535 was observed with REM using the $g'r'i'z'$ SDSS filters of the ROSS2 instrument (4000--9500$\AA$) between 2017 September 7th (MJD 58003) and October 1st (MJD 58027), obtaining strictly simultaneous images of the field. At the same time, thanks to the REMIR instrument, observations in the NIR ($J$, $H$, $K$ bands, 1.2--2.2 $\mu m$) were also acquired. For each night of observations, the single shots were averaged in each band, in order to enhance the signal to noise ratio. In this period of time, a total of 17 NIR-optical quasi-simultaneous data sets were obtained with REM. A detailed log of the observations can be found in Tab. \ref{REM_log}.
Each optical image was bias and flat-field corrected using standard procedures, and the magnitudes of all the objects in the field were extracted using the point spread function (PSF) photometry technique with the dedicated ESO-MIDAS\footnote{http://www.eso.org/projects/esomidas/}  {\tt daophot}\footnote{http://http://www.star.bris.ac.uk/$\sim$mbt/daophot/} task. The target was detected in the $z'$ and $i'$-bands only, in three and one epoch, respectively, probably due to the high extinction of the field ($E_{\rm B-V}=3.82$; see Sec. \ref{dereddening_sec} for more details). The flux calibration was performed against the standard star SA 110 \citep{Smith2005}.

NIR images were sky-subtracted, and the fluxes were extracted following the same procedure as in the optical. The NIR flux calibration was performed against a group of 12 isolated field stars in the 2MASS\footnote{http://www.ipac.caltech.edu/2mass/} catalogue \citep{Skrutskie2006}.

\begin{table}[htb]
\caption{Detailed log of the REM optical (ROSS2) and NIR (REMIR) observations performed between Sep. 7, 2017 and Oct. 1, 2017. }            
\label{REM_log}      
\centering                       
\begin{tabular}{c |c| c| c| c}       
\hline\hline                
Epoch & \multicolumn{2}{c|}{MJD mid observation} & \multicolumn{2}{c}{Exposures (s)} \\   
 (UT) &ROSS2 &REMIR &               $g'r'i'z'$&$JHK$           \\
\hline                     
  2017-09-07 & 58003.08438 & 58003.08538 & 3$\times180$  & 5$\times30$\\
  2017-09-08 & 58004.09137 &  58004.09236 &3$ \times180$  &  5$ \times30$\\
 2017-09-11 & 58007.06196 &  58007.06293 &3$\times180$  & 5$\times30$ \\
 2017-09-12 &58008.09050& 58008.09149 &3$\times180$ &  5$\times30$  \\
 2017-09-14 & 58010.08851 & 58010.08949 & 3$\times180$ &  5$\times30$\\
 2017-09-15 & 58011.09541&58011.09640 & 3$ \times180$ & 5$ \times30$ \\
 2017-09-16 & 58012.10244  &58012.10341 & 3$ \times180$  & 5$ \times30$ \\
 2017-09-21 &58017.02715 &58017.02814 &3$ \times180$ &5$ \times30$   \\
 2017-09-22 & 58018.04153& 58018.04252& 3$ \times180$ & 5$ \times30$  \\
2017-09-23 & 58019.04846& 58019.04944  &3$ \times180$  & 5$ \times30$ \\
2017-09-25 & 58021.04214& 58021.04312  &3$ \times180$  & 5$ \times30$  \\
2017-09-26 &58022.04907&58022.05006 &3$ \times180$ & 5$ \times30$   \\
2017-09-27 &58023.05597& 58023.05695 &3$ \times180$  & 5$ \times30$ \\
2017-09-28 &58024.07427 & 58024.07526  &3$ \times180$   & 5$ \times30$  \\
2017-09-29 &58025.08112 & 58025.08211  &3$ \times180$   & 5$ \times30$  \\
2017-09-30 &58026.08806 & 58026.08906  &3$ \times180$   & 5$ \times30$  \\
2017-10-01 &58027.09537 &  58027.09635 &3$ \times180$   & 5$ \times30$  \\
\hline                             
\end{tabular}
\end{table}

\subsection{Mid-infrared photometry with $VLT$ + $VISIR$}

Mid-IR observations of the field of J1535 were made with the Very Large Telescope (VLT) on seven nights from September 12th to 23rd 2017, under the program 099.D-0884 (PI: D. Russell). The VLT Imager and Spectrometer for the mid-Infrared \citep[VISIR;][]{LagageVISIR} instrument on the VLT's UT3 (Melipal) was used in small-field imaging mode (the pixel scale was 45 mas pixel$^{-1}$). Three filters ($M$, $J8.9$, $PAH2_2$) were used on different dates spanning central wavelengths 4.85--12.13 $\mu$m. The VISIR observing log, including photometric results, is reported in Tab. \ref{VISIR_res_tab}. For each observation, the integration time on source was 1000 s, composed of 22 nodding cycles. With chopping and nodding between source and sky, the total observing time was 1800--1900 s per observation.

The data were reduced using the VISIR pipeline in the \emph{gasgano} environment. Raw images from the chop/nod cycle were recombined and sensitivities were estimated based on standard star observations taken on the same night in the same filters. Aperture photometry was performed using an aperture large enough to ensure small seeing variations did not affect the fraction of flux in the aperture. The standard stars used are listed in the final column of Tab. \ref{VISIR_res_tab}. The flux densities of J1535 in the table were measured from the resulting combined frames.

We found that J1535 was bright enough on the first four epochs to measure the flux density of the source in individual nodding cycles. We produced short timescale light curves on these four dates; these are presented in Fig. \ref{variability_MIR_fig}. Clear MIR variations are observed on the time resolution of the observations, which is 80--90 s. In order to test whether these variations are intrinsic to the source, we investigate if sky variations, changing conditions or low S/N could account for the apparent variability. The overall long-term and night-to-night stability of the photometric calibration of VISIR is described in \cite{Dobrzycka12}, and the short-term stability was investigated during commissioning. From these tests, under photometric conditions the sky transparency variations were less than a few percent, and under clear conditions larger variability (up to $\sim 10$\%) were measured, although it is unlikely that the timescale of MIR variability would be as short as minutes, as appears to be the case for J1535. Moreover, the flux of J1535 varies by a factor of $\sim 2$--3 on timescales of minutes, which is in clear excess of ($\gtrsim 20$--30 times greater than) the variations expected from different conditions. VISIR has been used to study night-to-night variability in some objects \citep[e.g.][]{vanBoekel2010}, but we are not aware of any other published fast photometric variability studies using VISIR.

To further investigate if changing sky conditions could account for the apparent variability in J1535 at the time of the observations, we inspected the variations of the standard stars over the same dates. We found that the ADU/flux conversion factor measured from each standard star observation (in order to flux calibrate the data) varied by 7--8 \% over different dates (from September 11 to 23) under different weather conditions (clear or photometric) at different air masses. The variations in conversion factor on the level of 7--8 \% in the standards from night to night are likely due to differences in the sky conditions and airmass, as well as intrinsic differences between the conversion factors derived for different standard stars. J1535 clearly has lower S/N than the standard star observations, but the lower S/N cannot account for the observed variations in J1535 since the error bars on each flux measurement (which take into account variations in the background) are much smaller than the difference between fluxes in the same 30-minute light curves (Fig. \ref{variability_MIR_fig}). Possible background variations due to the water vapour content of the atmosphere above Paranal have moreover been monitored during our observations, revealing no clear variability of the background, that remained essentially constant during the observations. We therefore conclude that the MIR variations on minute timescales are almost certainly intrinsic to the source. In Tab. \ref{VISIR_res_var} we quantify the properties of the MIR variability of J1535 on the three dates in which the source was significantly detected in all 22 frames (these three dates were all under photometric conditions). We measured the intrinsic fractional rms variability (that subtracts the contribution to the variations from Poisson noise) adopting the method of \cite{Vaughan03,Gandhi2010}.

\subsection{Fast optical photometry with $SALT$}

J1535 was observed using the imaging camera $SALTICAM$ \citep{ODonoghue2006} on the Southern African Large Telescope ($SALT$) \citep{Buckley2006} near the beginning of the outburst, on Sept. 8th, 2017 at 17:37--18:34 UT (MJD 58004.7). 110 consecutive images of the field were made in the clear (white light) filter, each of exposure 30 s. These observations were done in frame transfer mode, with no deadtime between exposures. Automatic image reduction, star identification and relative magnitude measurements were run under the $PySALT$ pipeline \citep{Crawford2010}. J1535 was clearly detected in all frames. In Fig. \ref{variability_OPT_fig} we present the light curve, which shows intrinsic variability of J1535. The light curve of a nearby, slightly fainter field star is also shown, and is clearly much less variable. The variability properties are included in Tab. \ref{VISIR_res_var}, adopting the same method as above to calculate the fractional rms variability.
From differential measurements with USNO-B1.0 cataloged field stars, we estimated the brightness of MAXI at the time of the SALT observation to be $\sim$19.

\section{Results}

\subsection{Optical and infrared light curves: variability}\label{sec_variability}

\begin{figure*}
\centering
\includegraphics[scale=0.8]{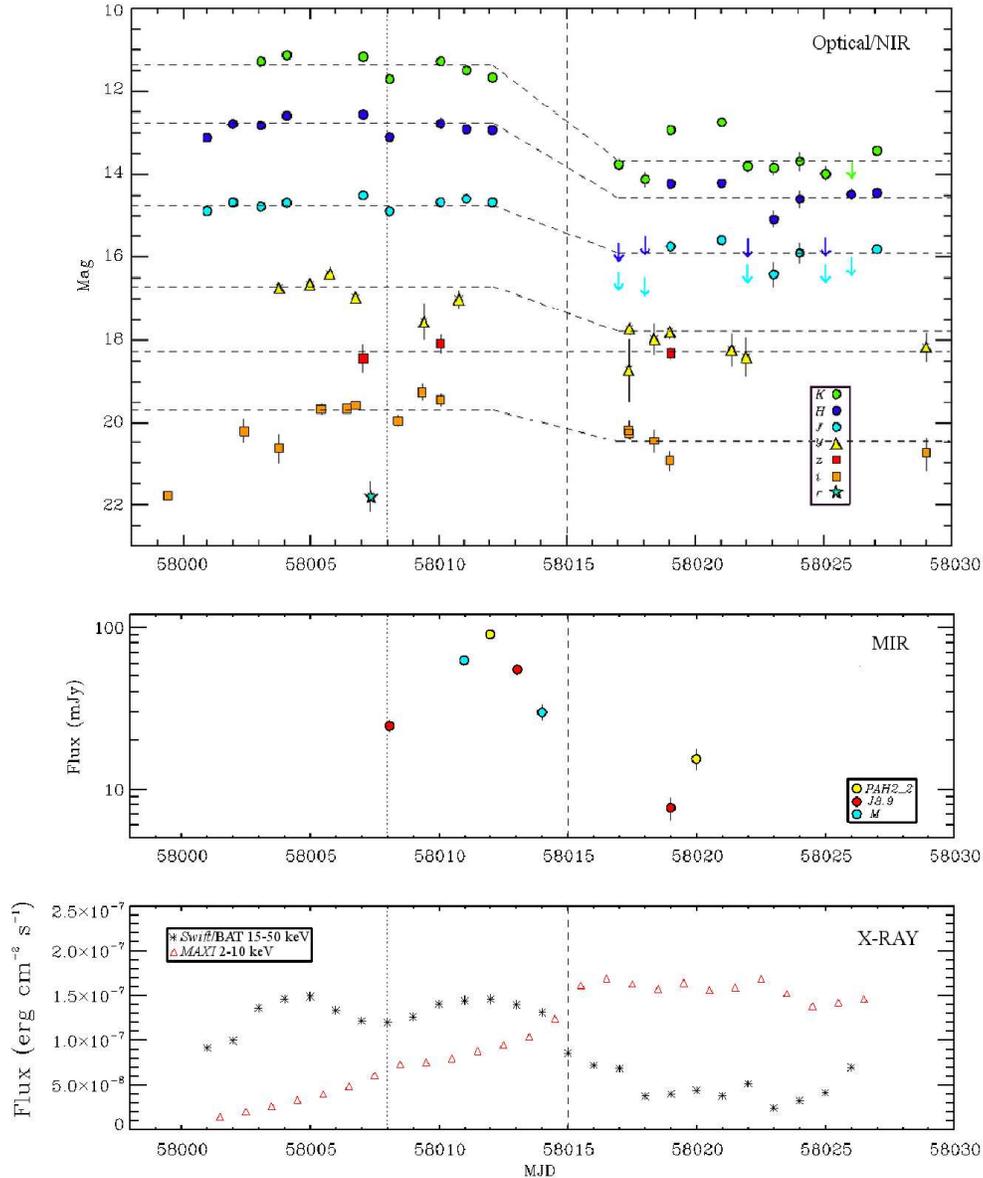}
\caption{\textit{Upper panel}: Optical and NIR light curves of J1535 during its 2017 outburst, obtained with the REM ($JHK$ and $i'z'$ bands) and the Faulkes ($r'i'y$ bands) telescopes from Sept. 6th to Oct. 3rd, 2017 (i.e. MJD 58002-58029). The SMARTS $JH$ detections reported by \citet{ATEL10716} and the $i'$ band detection reported by \citet{ATEL10702} are also shown (MJD 57999-58001). Errors are indicated at the $68\%$ confidence level. Magnitudes are not corrected for reddening. $3\sigma$ upper limits are indicated using arrows, where needed. Superimposed are the fits of the light curves with constant functions before and after the drop, to highlight the change. The two flaring points in the $JHK$-band, the upper limits and the first $i'$-band point have been excluded from the fits. \textit{Middle panel}: Mid-infrared (MIR) light curves of J1535 obtained with the VISIR instrument, using the $J8.9$, $M$ and $PAH2\_2$ filters. The values of the flux densities are not de-reddened. \textit{Bottom panel}: light curve of J1535 obtained with the BAT instrument mounted on \textit{Swift} (15--50 keV; black x), and with the \textit{MAXI} instrument (2--10 keV; red triangles) from Sept. 6th to Oct. 13th, 2017. Errors are indicated at the $68\%$ confidence level. The drop in the hard X-rays, corresponding to an increase of the soft X-ray flux, is clearly visible at MJD 58015 (dashed line in all panels) and the brief, slight softening is seen before that at MJD 58008 (dotted line in all panels).}
\label{OIR_lc}
\end{figure*}

\begin{figure}
\centering
\includegraphics[scale=0.4]{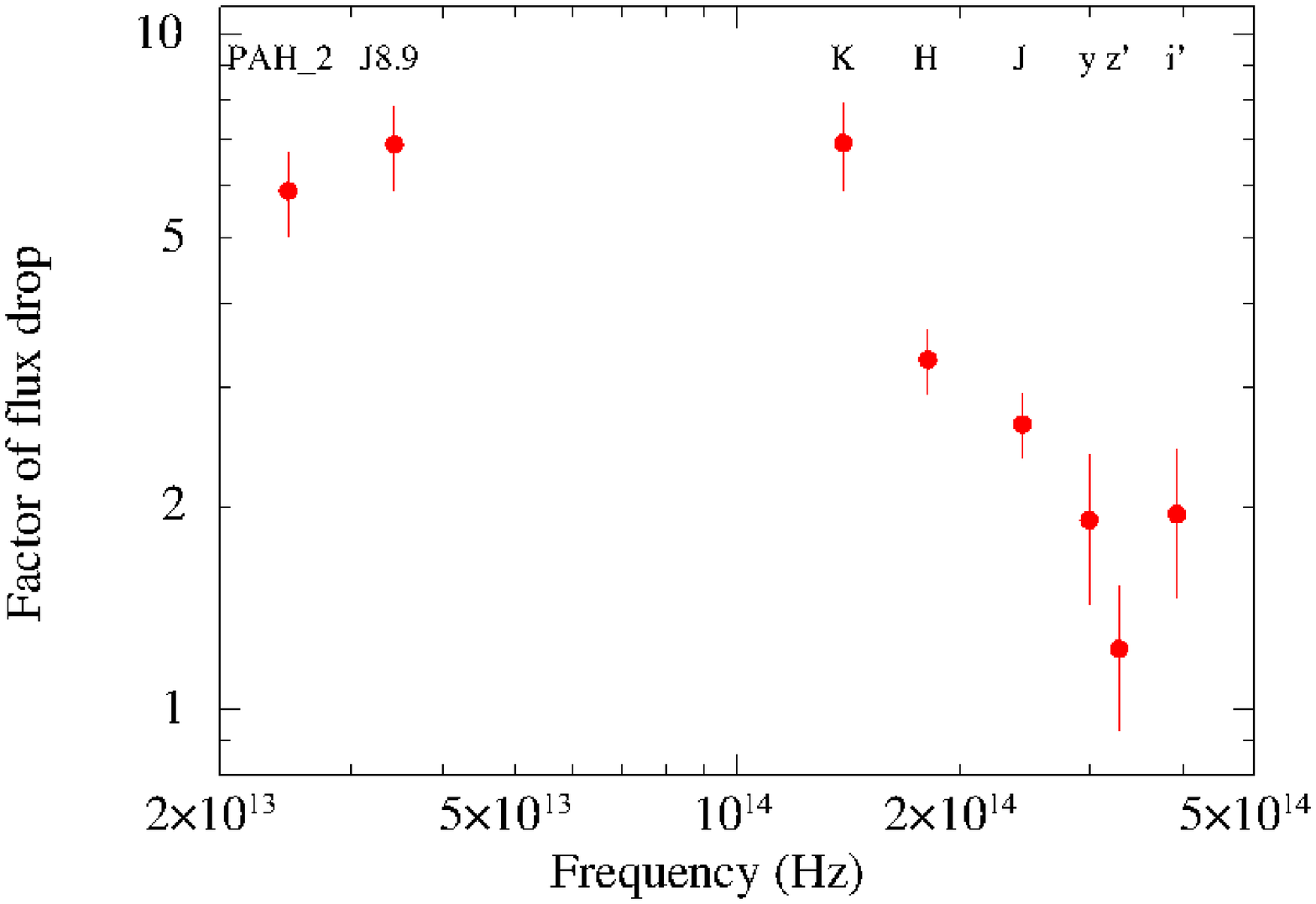}
\caption{Flux drop from MIR to optical as a function of the frequency. The drop is defined as the ratio between the flux soon before and soon after the transition to the soft state.}
\label{flux_drop}
\end{figure}

The optical and NIR light curves obtained with the REM and the LCO telescopes are shown in Fig. \ref{OIR_lc} (upper panel), in comparison with the MIR light curves obtained with VLT-VISIR (Fig. \ref{OIR_lc}, mid panel) and \textit{Swift} BAT 15--50 keV and \textit{MAXI} 2--10 keV (Fig. \ref{OIR_lc}, bottom panel) light curves during the same period of time. \\
The transition from a hard to a softer state reported in \citet{ATEL10761} on MJD 58015 is clearly visible in the \textit{Swift} light curve as a significant decrease of the hard X-ray flux. Unfortunately, we do not possess an optical/NIR observation on that day, which falls in a gap of our data-set; however, a large drop in the flux (more prominent going towards lower frequencies, reaching $\sim 2$ magnitudes in the $K$-band; see Fig. \ref{flux_drop}) is clearly noticeable between MJD 58012 and 58017, i.e. around the same time of the X-ray spectrum softening. This frequency-dependence of the fade suggests that, at least at NIR frequencies, the component that dominates the emission is likely to be the jet, that is quenched once the X-ray spectrum of the source softens, as expected in the case of BHBs (e.g. \citealt{Homan2005}; \citealt{Coriat2009}; \citealt{CadolleBel2011}; \citealt{Russell2013a}; \citealt{Russell2014}).
Despite the flux drop, the light curves display large variability on daily timescales, with both intra-observations flares and dips during the whole monitoring. In the NIR in particular, variations up to 1.2 mag in the $K$-band between two near epochs (i.e. on a timescale of $\sim 1$ day) are observed.
These features are uncommon for BH candidate LMXBs, that normally show a single NIR drop when the source transitions to the soft state when the jet is quenched, and then again a single NIR rise at the end of the outburst, when the source goes back to the hard state (see e.g. the case of GX 339--4, \citealt{Homan2005}; \citealt{Coriat2009}; \citealt{CadolleBel2011}; \citealt{Buxton2012}).


\begin{table*}[htb]
\caption{Observation log, and results of the mid-infrared photometry performed on J1535 with the VISIR instrument between Sep. 12 and Sep. 23, 2017. The reported fluxes are not de-reddened. Standard stars are: 1=HD198048; 2=HD161096; 3=HD163376; 4=HD000787; 5=HD156277; 6=HD211416; 7=HD178345; 8=HD133550.}            
\label{VISIR_res_tab}      
\centering                       
\begin{tabular}{c c c c c c c}       
\hline\hline                
Epoch (UT)                            & Filter & Wavelength & Weather & Airmass    & Flux density   & Standard\\   
 start time (MJD) &        & ($\mu$m)   &       conditions  &  & of J1535 (mJy) & stars observed\\
\hline                     
  2017-09-12 01:47:37 (58008.07474)  &$J8.9$  &8.72 & Clear	  & 2.04--2.35 & $25\pm 2$ & 1\\
  2017-09-14 23:19:22 (58010.97178)  &$M$     &4.85 & Photometric & 1.37--1.46 & $62\pm 5$ & 2, 3\\
  2017-09-15 23:29:54 (58011.97910)  &$PAH2\_2$&12.13& Photometric & 1.41--1.51 & $90\pm 2$ & 2, 3\\
  2017-09-17 00:53:36 (58013.03690)  &$J8.9$  &8.72 & Photometric & 1.78--2.01 & $55\pm 4$ & 1, 4, 5, 6\\
  2017-09-17 23:58:20 (58013.99884)  &$M$     &4.85 & Clear	  & 1.53--1.67 & $30\pm 3$ & 5, 7\\
  2017-09-22 23:41:34 (58018.98721)  &$J8.9$  &8.72 & Clear	  & 1.54--1.68 &  $8\pm 1$ & 8\\
  2017-09-23 23:42:32 (58019.98787)  &$PAH2\_2$&12.13& Clear	  & 1.56--1.70 & $15\pm 2$ & 8\\
\hline                             
\end{tabular}
\end{table*}

\begin{figure*}
\centering
\includegraphics[scale=0.3]{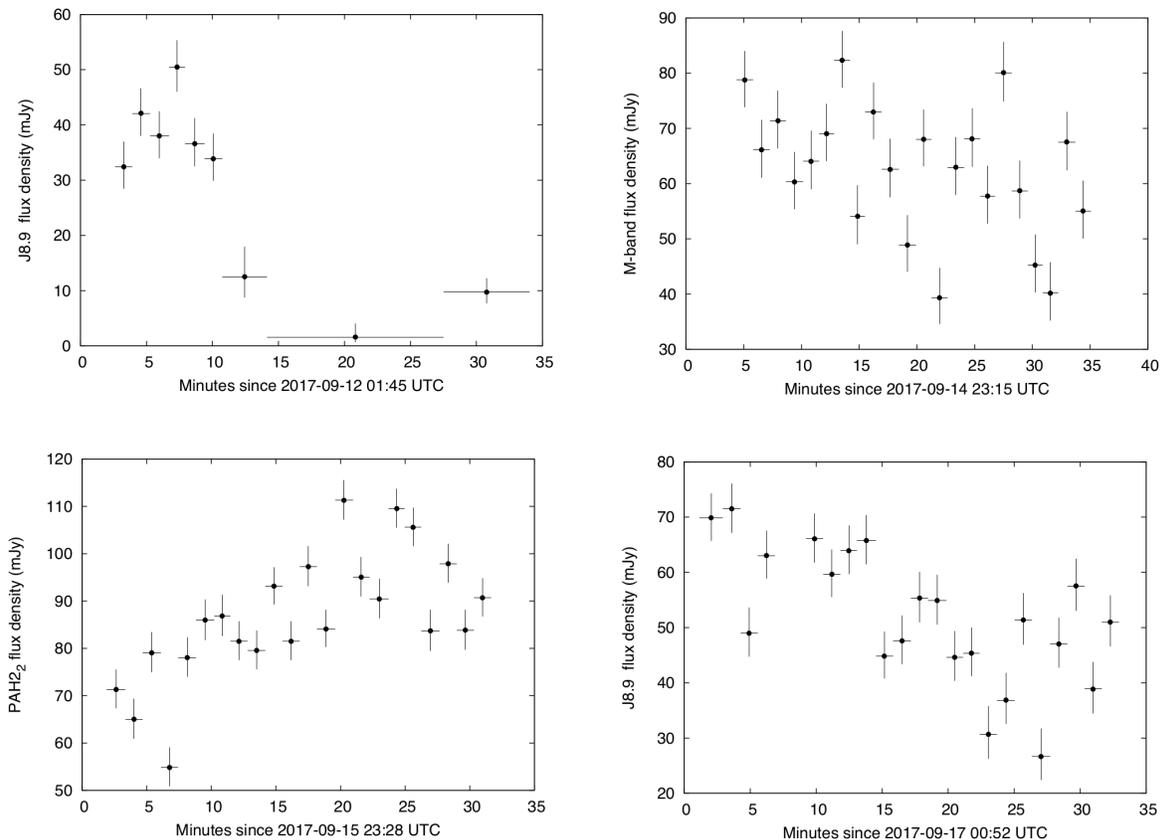}
\caption{MIR light curves of J1535 obtained with VLT--VISIR at four epochs: Sept. 12th (MJD 58008; $J8.9$ band), 14th (MJD 58010; $M$ band), 15th (MJD 58011; $PAH2\_2$ band) and 17th (MJD 58013; $J8.9$ band).}
\label{variability_MIR_fig}
\end{figure*}
\begin{figure}
\centering
\includegraphics[scale=0.15]{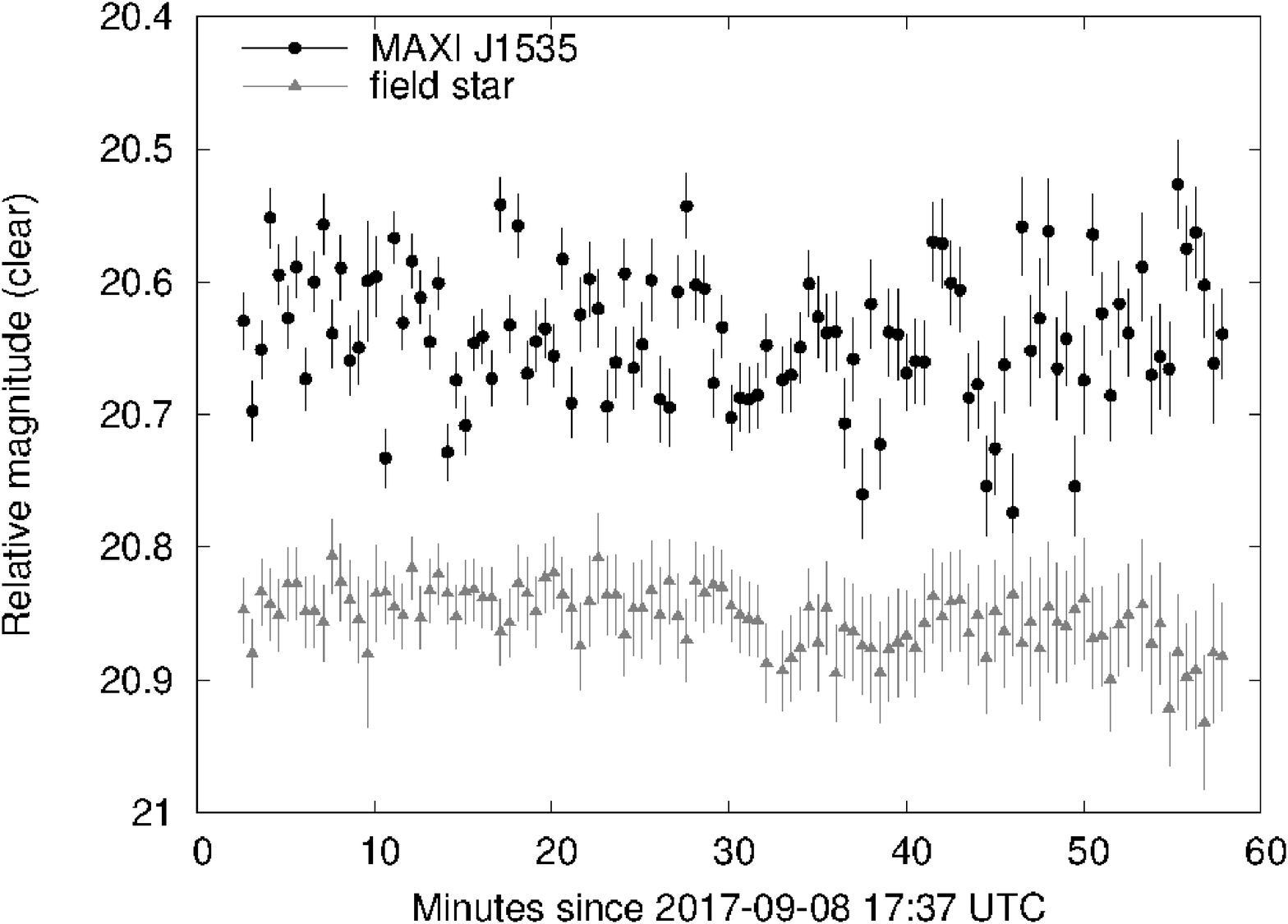}
\caption{Optical (white light) light curve of J1535 obtained with SALT--SALTICAM on Sept. 8th (MJD 58004.734).}
\label{variability_OPT_fig}
\end{figure}

Also in the MIR, a dramatic fading of the flux density corresponding to the softening of the X-ray spectrum around MJD 58015 is observed. In particular, a flux decrease of a factor of 7 in the $J8.9$ (8.72 $\mu$m) band between MJD 58008 and 58018 and of a factor of 6 in the $PAH2\_2$ (12.13 $\mu$m) band between MJD 58011 and 58019 is detected (see Fig. \ref{OIR_lc}, mid panel). \\
Moreover, a large amount of intrinsic variability is observed in the MIR (Fig. \ref{variability_MIR_fig}). The source being very bright on several epochs (see Tab. \ref{VISIR_res_tab}), the signal to noise ratio in the observations was high enough to search for short term ($\sim$ minutes) variations in the light curves obtained with the VISIR instrument. The brightest observation in particular (MJD 58011 in the $PAH2\_2$ band) can be split into 22 images of $\sim 80$s time resolution each, and the flux appears to vary significantly, by a factor of $\sim$2 in less than 15 minutes (see Fig. \ref{variability_MIR_fig}). The measured fractional rms is $\sim 15$--22 \% (see Tab. \ref{VISIR_res_var}), which is comparable to the fractional rms of GX 339-4 in the optical at a similar time resolution (\citealt{Gandhi2009}; \citealt{Gandhi2010}), and in the $K$-band ($13.20\pm0.05 \%$; \citealt{Vincentelli2018}). 

\begin{table*}[htb]
\caption{Optical and mid-IR photometric variability properties of J1535 with SALT and VLT (we include mid-IR data on the three dates in which the source was significantly detected in all 22 nodding cycles).}
\label{VISIR_res_var}      
\centering                       
\begin{tabular}{c c c c c c c c c}       
\hline\hline                
Epoch UT      & MJD &Filter   & Time           & Flux density     & Standard        & stdev/mean  & Fractional     & max/min\\   
 (Sept. 2017) & &         & resolution (s) & mean value (mJy) & deviation (mJy) & (\%)        & rms (\%)       & range factor\\
\hline                     
   8th        & 58004&white     & 30.4           & --               & --              &  4.7        &  $3.8 \pm 2.7$ & 1.26 \\
  14th        & 58010 &$M$       & 83.8           & 62.4             & 11.9            & 19.1        & $17.2 \pm 8.4$ & 2.09 \\
  15th        & 58011 &$PAH2\_2$ & 81.2           & 90.2             & 13.5            & 15.0        & $14.9 \pm 4.8$ & 2.03 \\
  17th        & 58013 &$J8.9$    & 87.1           & 54.8             & 12.3            & 22.4        & $22.0 \pm 8.7$ & 2.68 \\
\hline                             
\end{tabular}
\end{table*}


To investigate this variability further, we searched for short timescale variability also in the NIR dataset, by splitting the observations at higher S/N into five different images of 30s integration each. The NIR flux densities varied significantly on some of the dates, with a maximum variation of a factor of $\sim 2.5$ in less than 1 minute in the $H$-band on MJD 58008. Interestingly, MJD 58008 corresponds to the epoch before the softening of the X-ray spectrum in which the lowest flux was reported in our NIR campaign (see Fig. \ref{OIR_lc}), and shows light curves that are variable at a $3\sigma$ confidence level. These NIR observations are simultaneous to the last part of our MIR $J8.9$-band observations, which showed a sudden decrease of the flux density to undetectable levels after the first $\sim 10$ minutes of a high detected flux ($\sim 40 \,mJy$). 
Interestingly moreover, this temporary fading of the infrared emission corresponds to a dip in the hard X-ray light curve, i.e. to a short-lived softening of the X-ray spectrum.


\subsection{Dereddening}\label{dereddening_sec}

To obtain an estimate of the absorption column density and associated uncertainties, which is essential in order to de-redden our infrared and optical fluxes and to build broadband spectra of the target, we first started by fitting the $Swift$-XRT spectrum of J1535 acquired during a 1.1 ks observation carried out on 2017 September 11 (obs ID: 00010264004) with an absorbed power-law model. Pile-up effects were mitigated by excising the inner portion of the source point-spread function (within a radius of 12 arcsec). Absorption by the interstellar medium along the line of sight towards the source was modeled via the \textsc{TBabs} model \citep{Wilms2000} with abundances from \citet{Anders1989}. The fit was statistically acceptable, yielding a reduced chi squared $\chi^2_\nu = 1.08$ for 715 degrees of freedom. The inferred value for the column density and power law photon index were $N_H= (2.62\pm 0.02) \times 10^{22}$ cm$^{-2}$ and $\Gamma = 1.96 \pm 0.01$. This $N_H$ leads to a $V$-band absorption coefficient $A_V$ of $9.13\pm 0.50$ \citep{Foight16}.

If we consider the abundances from \citet{Wilms2000} instead of those of \citet{Anders1989}, we obtain a higher value of the $N_H$ ($3.84\pm0.03 \times 10^{22} \rm cm^{-2}$), that is consistent with what is reported in, e.g., \citet{ATEL10700}. With this $N_H$, $A_V=13.38\pm 0.73$ is inferred. The principal difference between the two abundances lies in the fact that \citet{Anders1989} uses the solar abundances as the reference ones for the intestellar medium (ISM), while \citet{Wilms2000} uses a more accurate estimate of the ISM abundances, even if still much uncertain (see \citealt{Wilms2000} for further details). \citet{Anders1989} abundances typically tend to favour a lower value of $N_H$, as we found.

The lowest value of the column density estimate for this source is instead given by the Galactic $N_H$, that is $1.43\times 10^{22}\, \rm cm^{-2}$ in the direction of the source \citep{Kalberla2005}, from which we estimate $A_V$ of $4.98\pm 0.27$, $E_{B-V}$ of $1.61\pm0.09$. This suggests that the $N_H$ measured from the modeling of the X-ray spectrum may have a strong component that is intrinsic to the source, and therefore may not be a good tracer of the dust absorption along the line of sight. 

Moreover, \citet{Tao2018} reported on evidence of variable $N_{H}$ during the days covered by our optical and infrared observations. However, if the local $N_H$ is not due to the presence of dust, the $N_H/A_V$ relation of \citet{Foight16} is not applicable, meaning that a changing $N_H$ does not necessarily affect the $A_V$ absorption. To verify if this is the case of MAXI J1535, we first considered an hypothetical source with constant infrared and optical fluxes, and we considered a possible changing $A_V$ due to the changing $N_H$ (as reported in \citealt{Tao2018}). We then de-reddened the constant fluxes of the source using these values of $A_V$, and we built new light curves. This test resulted in an optical $i'$ - band light curve which varied by 2.5 magnitudes (a factor $\sim$10 in flux), which is a higher amplitude than observed. This test also predicts the $K$-band light curve to be approximately constant but slightly brighter after the transition, whereas the opposite is observed (Fig. \ref{OIR_lc}).

We then searched for possible correlations between the infrared-optical colors measured for MAXI J1535 and the varying $N_H$ measured from the modeling of the \textit{Swift} spectra during the outburst \citep{Tao2018}, but no correlation was found. Therefore, we conclude that while the intrinsic $N_H$ is changing wildly (as measured from X-ray spectra), there is no corresponding expected change in the infrared-optical flux or colors, so intrinsic $A_V$ is negligible.

Further arguments that could explain a lack of dust intrinsic to the source might be that (a) the absorber resides in between the optical/IR and X-ray emission regions, and/or (b) the absorber is not large enough to cover the entire optical/IR emitting region, and/or (c) the absorber may be dust-free, or have a very low dust-to-gas ratio. \citet{Oates2018} recently discovered that in V404 Cyg (which had such high variable $N_H$ that it sometimes was Compton thick), the dust-to-gas ratio, must be of the order of $10^{4}$ smaller than that typically observed in the Milky Way, in order to not show significant change in color as the absorption increased by a factor of 1000. The absorber has to be within or at least close to the dust sublimation radius, in order to be dust-free. \citet{Oates2018} found that for V404 Cyg, the X-ray luminosity of V404 in quiescence resulted in a dust sublimation radius comparable to the size of the accretion disk. V404 Cyg is a long orbital period system, so it is extremely likely that for MAXI J1535 in outburst, with orders of magnitude higher X-ray luminosity than V404 Cyg in quiescence, the dust sublimation radius will be orders of magnitude outside the orbital separation of the X-ray binary. On the other hand, the intrinsic $N_H$ is likely to originate from an absorber located very close to the black hole (as was the case in V404 Cyg; see e.g. \citealt{Motta2017}), since it is variable on short timescales.

\subsection{Broadband spectral evolution}\label{spectrum_section}
The difficulty in determining the dust absorption coefficients in the direction of the source introduces large uncertainties when constructing the de-reddened OIR spectral energy distributions (SEDs) for J1535. Due to the high variability of the source, moreover, building an average spectral energy distribution of the target on a particular date is challenging. We therefore searched for as simultaneous as possible observations, in order to overcome at least this last issue. 

\begin{center}
\begin{figure*}

\includegraphics[scale=0.9]{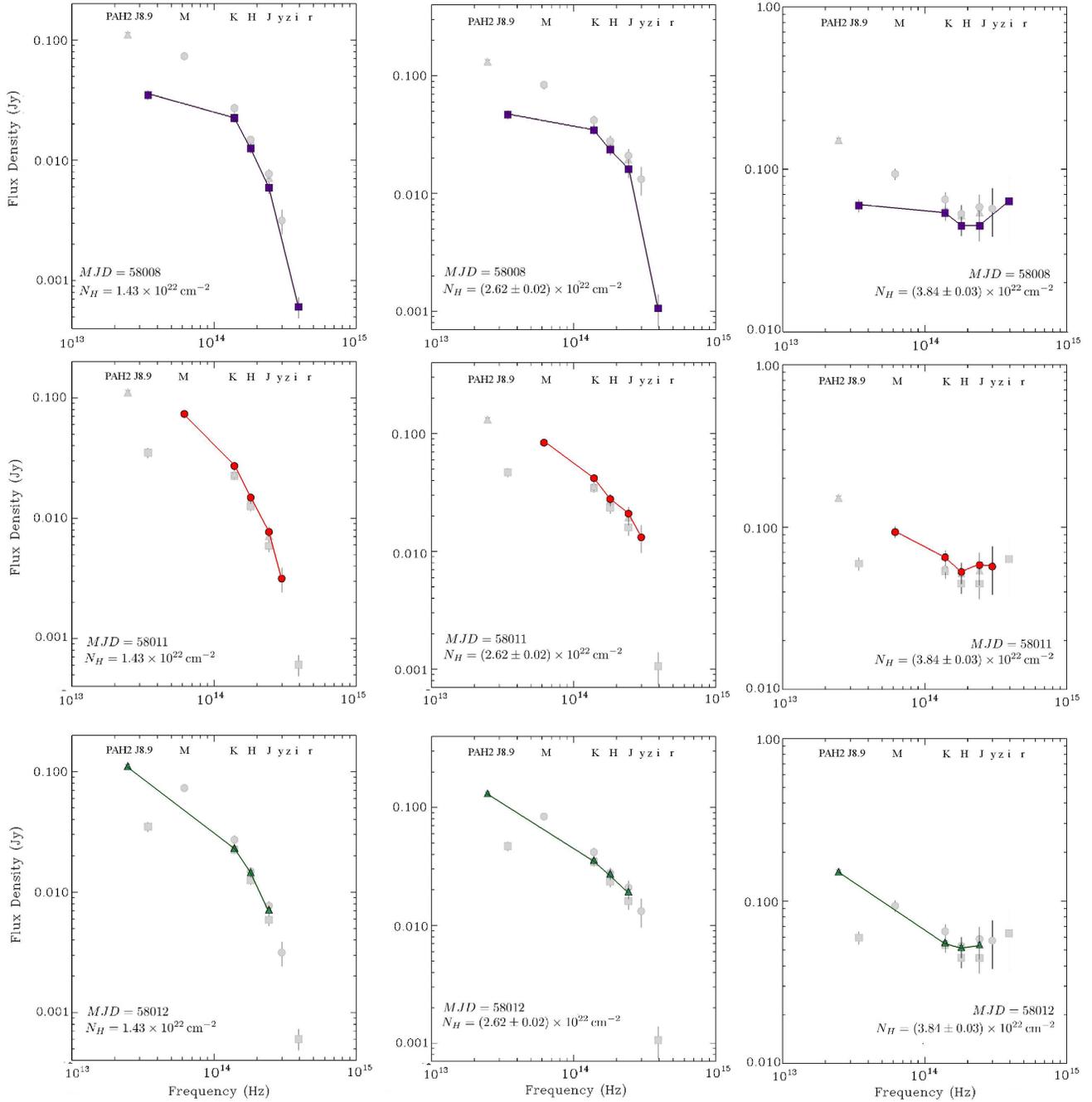}
\caption{Nearly simultaneous OIR spectral energy distributions (SEDs) of J1535 on MJD 58008 (purple squares), 58011 (red dots), 58012 (green triangles), obtained after de-reddening using the galactic $N_H$ ($1.43\times10^{22} \,\rm cm^{-2}$; left panel), $N_H=2.62\pm0.02\, \rm \times 10^{22} cm^{-2}$ (middle panel) and $N_H=3.84\pm0.03\,\rm \times 10^{22} cm^{-2}$ (right panel). In each panel, all the three SEDs are represented, for comparison purposes. Errors are shown at the 68$\%$ confidence level. }
\label{SED_08-11-12}

\end{figure*}
\end{center}

Thanks to the REM telescope, we possessed simultaneous data-sets in the optical, where unfortunately the target is barely detected only in the reddest bands due to the high extinction of the field. The NIR REM images are almost simultaneous with our optical observations (see Tab. \ref{REM_log}), and in three epochs (MJD 58008, -11, -12) we also have MIR observations that are very near in time to the REM ones (see Tab. \ref{VISIR_res_tab}; in particular, we notice that the last part of the MJD 58008 MIR observation overlaps with the NIR one). Therefore, we can construct the OIR SEDs at least on these three epochs, bearing in mind that the contribution of the jet might produce significant changes in the SED over minutes timescales, as observed in the MIR light curves. 
The three SEDs on MJD 58008, 58011, 58012 (2017 Sep. 12, 15, 16, respectively), de-reddened using three different values of $N_H$ ($1.43\times10^{22} \,\rm cm^{-2}$; $(2.62\pm0.02)\, \rm \times 10^{22} cm^{-2}$; $(3.84\pm0.03)\,\rm \times 10^{22} cm^{-2}$; see Sec. \ref{dereddening_sec}), are shown in Fig. \ref{SED_08-11-12}.

As is clearly visible in Fig. \ref{SED_08-11-12}, considerable changes in the shape of the SEDs are obtained by modifying the value of $N_H$ that is used to de-redden the fluxes. Using the lowest estimate of $N_H$, the NIR-optical spectrum is described by a power-law with a very steep slope ($\alpha \sim -3$), that is difficult to interpret. In fact, if we link the OIR emission of the system to the presence of a jet, we would expect to observe in the NIR-optical an optically thin synchrotron spectrum, which is normally described by a less steep power-law than what we observe (typically $\alpha \sim -0.8$; see e.g. \citealt{Gandhi11}). Instead, if the accretion disk is contributing to the NIR-optical emission, we would expect to observe a positive slope, due to the black body tail of the disk, which is clearly not our case. This is therefore an indication that the galactic $N_H$ is probably an underestimate of the real absorption of light on the target line of sight. 

The second estimate of $N_H$ ($2.62\pm0.02\, \rm \times 10^{22} cm^{-2}$; Fig. \ref{SED_08-11-12}, mid panel) gives instead not extremely steep SEDs (except for MJD 58008, where the $i$- band flux is really low with respect to the lower frequency points), with a slope of  $-\sim 1.5$ and $-0.8$ on MJD 58011 and 58012, respectively. However, the contribution of the accretion disc is still missing at the highest frequencies, which is uncommon for an X-ray binary in outburst.

Instead, using $N_H=3.84\pm0.03\,\rm \times 10^{22} cm^{-2}$, the contribution of the disk starts to be evident, and a clear infrared excess due to the emission of a jet is visible (see Fig. \ref{SED_08-11-12}, right panels). In particular, the slope of the power-law that we used to fit the lower frequency points of the MJD 58011-12 SEDs are $\sim -0.8$ - $-0.6$, respectively, which are values that are typically measured for the optically thin synchrotron spectrum of a jet.

For this reason, from now on, the value of $N_H$ of $3.84\pm0.03\, \rm \times 10^{22} cm^{-2}$ will be used to de-redden the fluxes of J1535. The corresponding absorption coefficients evaluated for all wavelengths that are relevant in this work are reported in Tab. \ref{tab_absorption}. However, we caution the reader that the de-reddened spectral slopes depend critically on the uncertain extinction.

\begin{table}[htb]
\caption{Absorption coefficients derived from $N_H=3.84 \pm 0.03 \times 10^{22} \,\rm cm^{-2}$, and using the relation between $N_{H} $ and $A_{V}$ reported in \citet{Foight16}, and the coefficients reported in \citet{Schlafly2011} and \citet{Weingartner2001} for the NIR-optical and the mid-infrared wavelengths, respectively.}            
\label{tab_absorption}      
\centering                       
\begin{tabular}{c c c}       
\hline\hline                
Filter & Wavelength & $A_{\lambda}$\\
      &    $(\mu m)$    &  (mag) \\ 
\hline                     
$PAH2\_2$ & 12.13 & $0.60\pm 0.02$ \\
$J8.9$      & 8.72 &$1.03\pm0.03$ \\
$M$  &   4.85 &$0.47 \pm 0.02$ \\
$K$ & 2.16 & $1.51\pm 0.08$ \\
$H$ &1.66 & $2.20 \pm 0.12$ \\
$J$ & 1.23 &$3.51\pm 0.19$ \\
$y$ & 1.00 &$5.01\pm0.27$ \\
$z'$ & 0.91 & $5.91\pm0.32$ \\
$i' $& 0.76 & $8.05\pm0.44$ \\
$r'$ & 0.62 &$10.96\pm0.60$ \\
\hline                             
\end{tabular}
\end{table}

Focusing on the right panel of Fig. \ref{SED_08-11-12}, we notice that on all dates the negative index power law extends up to the MIR. If we interpret it as optically thin synchrotron emission from the jet, this suggests that the jet break frequency might fall at lower frequencies than the MIR, i.e. in the radio to far-IR bands (with upper limits to the jet break frequency of $6.2\times10^{13}$ Hz and $2.5\times10^{13}$ Hz on MJD 58011 and 58012, respectively).
On MJD 58008 the slope of the SED at lower frequencies is lower with respect to MJD 58011 and 58012, which suggests for a lower contribution of the jet at those frequencies on that day. However, we caution that MJD 58008 is the epoch in which the mid-infrared observation shows the sudden drop in flux after the first $\sim 15$ mins of observation, until it becomes undetectable. This strong, short-timescale MIR variability might therefore distort the low-frequency shape of the SED at this epoch. Moreover, \cite{ATEL10745} reported on radio detections of J1535 with ALMA on the same day, with a preliminary flux measurement of $\sim220$ mJy at 7 and 140 GHz, which is much higher than what we measure in our $J8.9$-band VISIR observation ($\sim 63 $ mJy after de-reddening). 

\begin{figure}
\centering
\includegraphics[scale=0.47]{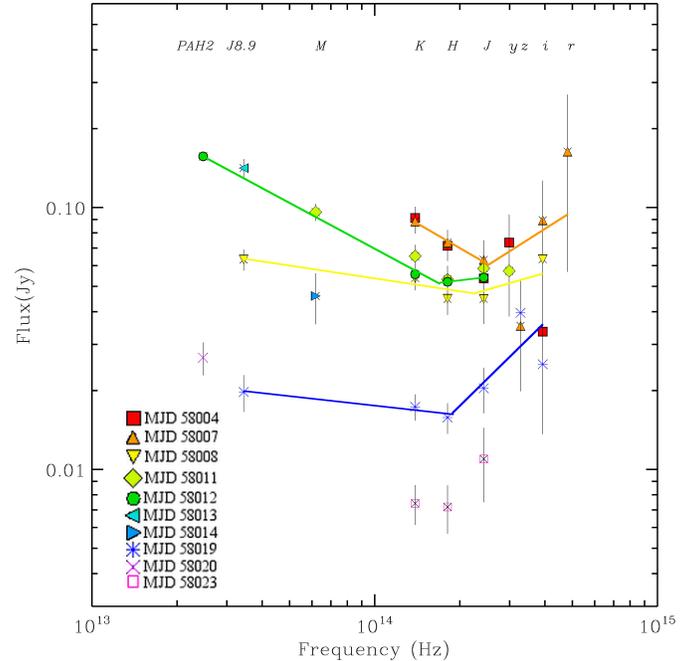}
\caption{Spectral energy distributions of J1535 from MJD 58004 to MJD 58023. The SEDs evolve from red (MJD 58004) to purple (MJD 58023). The flux densities at all wavelengths have been de-reddened using the absorption coefficients reported in Tab. \ref{tab_absorption}. Errors are reported at the 68$\%$ confidence level. }
\label{SED_tot}
\end{figure}

In Fig. \ref{SED_tot}, all the nearly simultaneous SEDs collected in our campaign are shown, from red to purple following the rainbow colors (as explained above, $N_H=(3.84\pm0.03)\times 10^{22}\,\rm cm^{-2}$ has been used to de-redden the fluxes). The drop in the NIR flux that is observed in the light curves (see Fig. \ref{OIR_lc}) after MJD 58012 is clearly noticeable in the SEDs, with a large modification of the NIR SED shape between MJD 58012 and MJD 58019. In particular, if we fit the NIR-optical SEDs with a power-law, the slope evolves from almost flat to positive (see Fig. \ref{SED_tot}), with a residual infrared flux in the MIR, which might testify that the contribution of the jet is lower, but still detectable at the lowest frequencies, even if the X-ray spectrum softened (as reported in \citealt{ATEL10761}). Also in the optical, the small drop in flux due to the decrease of the jet contribution is visible in the SEDs. A small drop is expected if the accretion disk has a stronger contribution to the optical than the NIR.

If we then look at the MIR part of the SEDs, we can observe the fluxes changing considerably between the different epochs. The highest average flux is reached in the PAH$2\_2$ band (12.13 $\mu m$), on MJD 58012, with a de-reddened value of $157\pm4$ mJy; after that, the average MIR flux decreases, reaching its lowest value on MJD 58020 ($27\pm4$ mJy). The first measurement after the softening of the X-ray spectrum is a de-reddened flux density in the $J8.9$- band of $20\pm3$ mJy (MJD 58019), that is a factor of 7 lower than the flux density measured soon before the softening in the same band (i.e. $141\pm12$ mJy on MJD 58013). It is therefore evident that we are observing in the MIR the emission of a transient component, that is suppressed as the X-ray spectrum softens, as usually happens in case of jets quenching over the hard to soft transition in BHBs.

\section{Hardness Intensity Diagram}\label{HID_section}
\begin{figure*}
\centering
\includegraphics[scale=0.6]{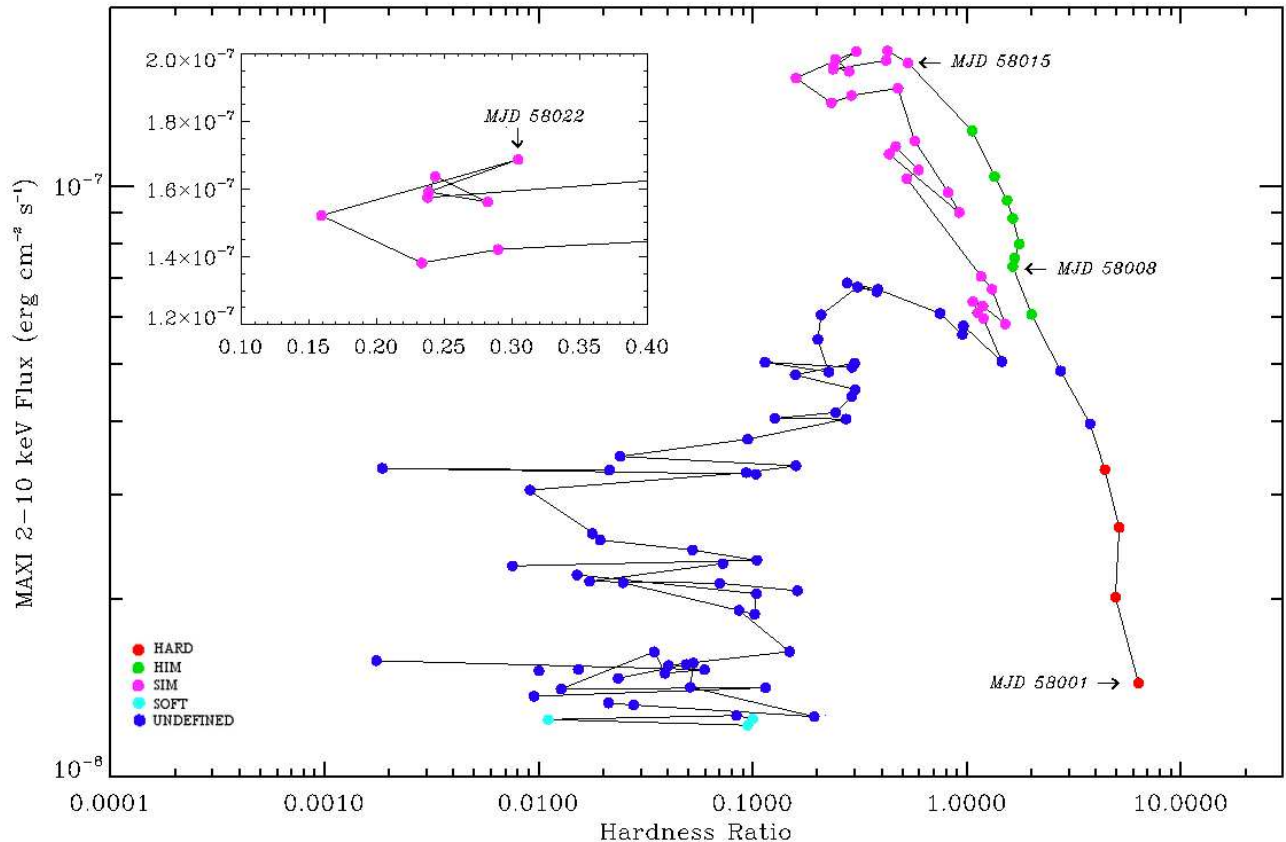}
\caption{Hardness-intensity diagram (HID) for the 2017-8 outburst of J1535. The color-code follows the state definition of \citet{Tao2018}, according to which four principal states can be observed: the hard state (red dots), the hard-intermediate state (green dots), the soft-intermediate state (violet dots) and the soft state (light blue dots). In the figure, the first softening of the X-ray spectrum on MJD 58015, which corresponds to the drop in the OIR fluxes that we report in this work, is highlighted. The inset shows instead the sudden increase in the hardness on MJD 58022, which might correspond to the flares that we detect in the NIR light curves.}
\label{HR_fig}
\end{figure*}

We constructed the HID of J1535 (Fig. \ref{HR_fig}) in order to study the evolution of the X-ray spectrum of the target during the first part of its outburst (from MJD 58001 to MJD 58139, i.e. Sep. 5th, 2017 - Jan. 21st, 2018), and to test whether the IR flaring episodes correspond to changes in the evolution in the HID.
We considered the \textit{MAXI} 2--10 keV count rate obtained almost daily during the monitoring of the outburst as an estimate of the soft X-ray photons, and the \textit{Swift} BAT (15--50 keV) count rate for the hard ones, and we evaluated the X-ray color (or hardness ratio, HR) as the ratio between the two. The count rates were converted into flux measurements using the WebPIMMS\footnote{https://heasarc.gsfc.nasa.gov/cgi-bin/Tools/w3pimms/w3pimms.pl} tool. 

During the rise of the outburst, until MJD 58015, the source shows a hard X-ray spectrum, and, as already pointed out in this discussion, very bright radio detections of flat/inverted spectrum jets have been reported (\citealt{ATEL10711}; \citealt{ATEL10745}). These are typical features of BHBs in outburst, and are all accounted for in the \citet{FenderBelloniGallo04} model for the accretion-ejection coupling mechanism. Following the X-ray state classification reported in \citet{Tao2018} in particular, MAXI J1535 underwent a first state change from the hard state (red dots in Fig. \ref{HR_fig}) to a hard-intermediate state (HIM; green dots in Fig. \ref{HR_fig}) between MJD 58004 and 58007. This smooth change of the spectrum hardness is also visible in Fig. \ref{HR_fig}.
At MJD 58015, the spectrum undergoes a first softening (as reported in \citealt{ATEL10761}). The same softening of the X-ray spectrum is also reported in \citet{Tao2018} using \textit{Swift} observations. According to them, on MJD 58015 the source entered the so-called soft-intermediate state (SIM; violet dots in Fig. \ref{HR_fig}), where it remained until MJD 58050, i.e. for the whole duration of our infrared-optical campaign. According to the accretion-ejection coupling scenario, this intermediate phase should correspond to the NIR fading, as observed in other BHB candidates when they start to soften (see e.g. GX 339-4; \citealt{Homan2005}; \citealt{Coriat2009}; \citealt{CadolleBel2011}). This is exactly what we observe for J1535.

Then, just four days after the softening, on MJD 58019 we detect the already mentioned flickering activity in the NIR light curves, with a brightening by more than 1 mag (i.e an increase of the flux by a factor of 2.5) on a 1-day timescale in the $K$-band (and the source became visible again in the $J$ and $H$ bands after the drop). The enhanced activity in the NIR lasted at least two days, until MJD 58021. In the HID shown in Fig. \ref{HR_fig}, we notice that this few days period corresponds to a change in the hardness of the system, with a temporary inversion of the decreasing trend of the hardness (see the inset in Fig. \ref{HR_fig}). After MJD 58022, the hardness started again to decrease. Similar excursions of the hardness are also observed later, up to $\sim$MJD 58051, i.e. when the source finally entered its soft state. Unfortunately we do not possess a NIR coverage with REM after October 1st (i.e. MJD 58027) to show a possible NIR counterpart to these hardness changes, since the source was no longer visible from La Silla. J1535 did not make a smooth transition to the soft state. The excursions in the HID starting around MJD $\sim 58018$ appear to be associated with IR variations from an intermittent jet.

\section{Discussion}
In this paper we present an analysis of the optical-infrared (MIR and NIR) light curves and spectral energy distributions of J1535 during the first part of its 2017-2018 outburst. 
\subsection{Jet suppression after the X-ray spectrum softening }
As shown in Fig. \ref{OIR_lc}, the NIR and optical fluxes stay at a high level during the first phases, which correspond to the rise of the X-ray outburst; at the same time, the X-ray spectrum was reported to be hard by several authors (see Sec. \ref{sec_maxi}), as typically happens during the rise of a BHB X-ray outburst. After MJD 58015 however, i.e. when $Swift$ detected a first softening of the spectrum \citep{ATEL10761}, the infrared and optical fluxes underwent a prominent drop, which indicates that one of the players in the infrared-optical emission was experiencing a suppression in correspondence to the softening of the X-ray spectrum. 

From studies of the coupling between accretion and ejection in BHBs it is clear that a compact jet is produced during the hard state of BHB outbursts, that is then quenched as the system passes to the soft state, due to the observation of a strong suppression of the radio emission after the transition (see \citealt{Tananbaum1972}; \citealt{Fender1999}; \citealt{FenderBelloniGallo04}; \citealt{Corbel2004}). Less clear is whether the same effect should also be visible at higher frequencies, where the contribution of the jet should not always dominate the emission. However, several cases in which this happens were observed so far (see e.g. \citealt{Homan2005}; \citealt{Russell2006}; \citealt{Coriat2009}; \citealt{Russell2013b}). In particular, if the jet largely contributes to the IR-optical emission, it is likely that we will observe a weakening of the IR-optical flux during a hard-to-soft state transition, which should be more prominent towards lower frequencies, where the contributions of the outer accretion disk and of the X-ray reprocessing are typically lower. This is exactly what we observe: the flux drop corresponding to the softening of the X-ray spectrum is in fact higher in the MIR and in the NIR, where we have a decrease of a factor of 6 at 12.13 $\mu$m on an 8 day time scale, than in the optical, where the measured drop corresponds to 0.91 mag in the $i'$-band (i.e. a factor of $\sim$2 in flux. See Fig. \ref{flux_drop}).

If we follow the jet suppression interpretation to explain the flux drop that we observe, this implies that the jet was contributing at least $83 \pm 4 \%$ of the flux in the PAH2$\_2$ band (12.13 $\mu$m), $86 \pm 9 \%$ in the $K$-band, and $57 \pm 21 \%$ in the $i'$-band. This is consistent with what was found in \citet{Russell2006} for a sample of BHBs, for which the jets have been found to contribute $\sim90\%$ of the NIR emission and below $76\%$ of the optical emission in the hard state.

Inspecting the SEDs (Fig. \ref{SED_tot}), we see that before the softening occurred on MJD 58015 in the IR regime the spectrum is described by a power law with a negative index, with slopes ranging from -1 to -0.6, that are typical of optically thin synchrotron emission from a jet (adopting $N_{H}=(3.84\pm0.03)\times 10^{22}\,\rm cm^{-2}$). In the optical, the slope of the power-law is always positive, but it becomes steeper after the drop, which testifies that the accretion disc contribution becomes more important with respect to the emission of the jet at those frequencies after the drop. This is also a further evidence of the fact that the jet is contributing not only in the radio band, as reported by \citet{ATEL10711} and \citet{ATEL10745}, but also at higher frequencies, up to the optical band, with the jet break frequency probably lying at lower frequencies than the MIR. We therefore interpret the observed drop in the IR-optical flux as the weakening of the emission of the jet while the X-ray spectrum softens.

\subsection{The flickering nature of the jet in J1535}
The fade of the IR--optical flux near the start of the transition towards the soft state has already been observed in other BHBs. After the drop, the IR flux usually does not increase again, until the source transitions back to the hard state in the last phases of the outburst. Indeed, this was the case for XTE J1550-564 (\citealt{Jain2001}, \citealt{Russell2010}), MAXI J1836-194 (\citealt{Russell2013b}; \citealt{Russell2014}), and from repeated outbursts of GX 339-4 (\citealt{Coriat2009}, \citealt{Buxton2012}, \citealt{Dincer2012}; see also \citealt{Kalemci2013}, \citealt{Corbel2013}). However here, we find that after the softening in J1535, the optically thin component seems to be still present in the MIR, even if strongly suppressed. In addition, inspecting the infrared light curves (Fig. \ref{OIR_lc}), the NIR flux experiences a sudden increase on MJD 58019-21, which does not have any correspondence with the optical light curves. This excludes that this flaring NIR activity might originate in the reprocessing of the accretion disk emission. It is therefore likely that the origin of these NIR flares is in the jet, which might experience a re-flaring after the target started to soften. This hypothesis is supported by the SEDs, where it is clear that after the softening the jet seems to be still present at lower frequencies (MIR). Nearly simultaneous observations in the radio band might confirm or confute this interpretation.

As described in the semi-quantitative model of \citet{FenderBelloniGallo04} and \citet{Fender2009} (see also \citealt{Vadawale2003}), when a BHB is in a transitional phase between the hard and the soft state it is likely to produce several synchrotron spectra from discrete ejections, which evolve towards lower frequencies with time, as the compact jet becomes more intermittent. This is seen most prominently at radio frequencies, later during the transition, close to the `jet line' when bright, optically thin radio flares are commonly observed. Multiple radio flares have been observed when BHBs perform hard--soft excursions in the hardness-intensity diagram (HID) during intermediate states. For example, GRS 1915+105 undergoes small loops in its HID which are accompanied by many radio flares from multiple ejections \citep{FenderBelloniGallo04}. Here, we instead detect IR flaring at earlier (harder) stages of the spectral transition.


In addition to the flaring after the main IR fading, we also notice that on MJD 58008 (hours before our infrared observations) \textit{ALMA} mm and \textit{ATCA} radio observations have been reported (see \citealt{ATEL10745}; Sec. \ref{sec_maxi}), revealing that the flux in the millimeter band is found at a much higher value than what we measure in the MIR ($>200$ mJy at 97 and 140 GHz, vs. $\sim 63$ mJy in the $J8.9$ band of VISIR after de-reddening). As described in Sec. \ref{sec_variability}, the MIR light curve that we extracted for the MJD 58008 epoch is highly variable, the source becoming undetectable after a certain time. Interestingly, the same decrease in the flux is also observed in the NIR on the same date, which on that day is strictly contemporaneous with the last part of the MIR observations (i.e. when the flux has already decreased). This behavior could again be interpreted as an intermittent jet, as well as flares within the jets, and appears to be associated with a small excursion in the HID, before the main softening. 

\subsection{Short time variability: possible interpretation}
J1535 is the first BHB for which mid-infrared short time variability has been observed so far (with rms $\sim 15$--22\%; Tab. \ref{VISIR_res_var}). It is suggestive for the presence of a flickering jet, similar to what was found from mid-IR observations of GX 339-4 on longer (hour) time scales \citep{Gandhi11}. Similar levels of fractional rms were observed in other BHBs at higher frequencies, like in the case of GX 339--4 in the optical \citep{Gandhi2010}. In that case, a study of the correlation between optical and X-ray (fast and slow) variability allowed the authors to exclude the accretion disk radiation reprocessing as the cause of the observed $\sim 15\%$ optical rms, which was moreover higher for redder frequencies. Similarly to what was suggested by \citet{Gandhi2010}, a model that relates possible perturbations in the accretion flow to variability in the jet might explain the observed variability of our light curves. Such a model has been developed in \citet{Malzac2013} and \citet{Malzac2014} \citep[see also][]{Drappeau2017}. In this scenario, a strongly variable accretion flow could inject non-negligible velocity fluctuations at the base of the jet, that would drive internal shocks at large distances from the BH, where leptons are accelerated, giving rise to strongly variable synchrotron emission, which would in turn affect the radio/infrared emission of the system. This model has been found to be compatible with the mid-IR variability and SED observed by \citet{Gandhi11} in GX 339-4 \citep{Drappeau2017}.
We have found that the strong observed infrared variability is correlated with changes in the X-ray emission of the system on long timescales (days).

A detailed quantitative comparison of the consistency of the internal shocks model (\citealt{Malzac2013}, \citealt{Malzac2014}) with our observations goes beyond the scope of this paper. However, a comparison between what we observe and the expectations from the model can be easily performed. Using fig. 6 of \citet{Malzac2014}, we can infer the value of the fractional rms that is expected for optical and infrared light curves in a BHB with typical jet parameters (average Lorentz factor $\gamma=2$ for a kinetic power $P=1.3\times 10^{37} \rm erg\, \rm s^{-1}$, jet half opening angle $\Phi=1^{\circ}$, mass of the black hole $M=10M_{\odot}$, etc; see Sec. 4 of \citealt{Malzac2014} for more details). In particular, it is clear from the figure that the expected optical rms should be higher than the mid-infrared one in the Fourier frequency range that spans from the frequency corresponding to the total duration of each of our MIR light curves ($\sim 30$ min, out of scale in Fig. 6 of \citealt{Malzac2014}) to that associated to the time resolution ($\sim$ 85 s; see Tab. \ref{VISIR_res_var}). In the MIR, the fractional rms predicted by the model is $\sim 17\%$ at a Fourier frequency of $1.2 \times 10^{-2}$ Hz, which corresponds to the time resolution of our MIR observation for which the most significant rms has been detected (81.2 s with the $PAH2\_2$ band on Sept. 15th; see Tab. \ref{VISIR_res_var}). Interestingly, this value is consistent with our findings (rms $\sim 15\%$; Tab. \ref{VISIR_res_var}).
In the optical instead, we obtain from the SALT light curve a much lower rms than expected (only $3-4\%$ rms, see Tab. \ref{VISIR_res_var}). However, since J1535 is a BHB in outburst, it is likely that the accretion disk is playing a major role in the optical, thus diluting the jet emission at those frequencies, and therefore decreasing the detected fractional rms due to the internal shocks in the jet.
A more detailed calculation of the expected rms should be performed by tailoring a simulation for J1535, so that it can reproduce the SED of the target. However, we do not expect the predicted rms to vary dramatically with the model parameters.

\section{Conclusions}

In this work, we presented optical and infrared observations of the newly discovered black hole candidate transient X-ray binary MAXI J1535--571 during the first phases of its recent outburst, which started on MJD 57998 (Sept. 2nd, 2017).

During the rise of the outburst, X-ray observations revealed a hard spectrum, as typical of BHBs, while the NIR and optical fluxes remained high. During this period of time, until MJD 58015 (Sept. 19th), the infrared spectral energy distribution was consistent with a power law with a negative index, likely due to the presence of a jet, as observed in the radio band. On MJD 58015, \textit{Swift} detected a softening of the spectrum, which coincided with a drop of the infrared and optical fluxes of a factor of 10 and 2, respectively, on a timescale of a few days. The drop was more significant at lower frequencies, and we interpreted it as due to the suppression of the emission of the jet, that is more dominant in the infrared than in the optical. A further confirmation to this scenario came from our mid-infrared campaign, performed with the VISIR instrument on the VLT. In fact, also the mid-infrared fluxes experienced a significant drop (of a factor of $\geq 6$) in correspondence to the observed softening of the X-ray spectrum, thus reinforcing our hypothesis.
 
The mid-infrared light curves have been found to possess a significant variability over a timescale of a few minutes, with a maximum measured rms of the $\sim 15$--22\%. This makes MAXI J1535--571 the first BHB for which a mid-infrared short timescale variability study has been performed so far. This strong variability suggests the presence of a flickering jet, as previously observed in other BHBs using NIR--optical data \citep{Casella2010,Kalamkar2016,Shahbaz2016,Gandhi2016,Gandhi2017,Vincentelli2018}. The measured values of rms in the mid-infrared might be consistent with the internal shock model developed for jets in X-ray binaries (\citealt{Malzac2013}, \citealt{Malzac2014}). We also report a re-flaring in the near-infrared light curves, after the main suppression of the jet. These late synchrotron flares are associated with X-ray hardness variations, which may be consistent with a jet spectral break frequency correlating with the X-ray hardness \citep{Koljonen2015}.

A comparison of our results with lower frequency data (i.e. the radio band) is essential in order to clarify the proposed scenario, and will be part of a follow-up paper on the outburst of this source.

The observations of rapid mid-IR flickering in J1535 opens up a new field: studying the fast mid-IR variability in X-ray binaries. Such studies performed at optical and NIR wavelengths, especially conducted simultaneously with X-ray observations \citep[e.g.][]{Gandhi2010,Casella2010,Durant11,LassoCabrera13,Kalamkar2016,Gandhi2017}, have led to leaps forward in our understanding of disk/jet physics. However, with current mid-IR instrumentation, it is not possible to observe BHBs (or neutron star LMXBs) at time resolutions less than minutes with high enough S/N due to sensitivity limitations. With the advent of NASA's James Webb Space Telescope (JWST; see \citealt{Kalirai2018} for an up-to-date review), MIR sensitivities will be increased by orders of magnitude compared to current ground-based facilities like VISIR on the VLT. As such, it will be possible in the future to measure sub-second variability of LMXBs at MIR and NIR wavelengths for the first time with JWST.

\begin{acknowledgements}
DMR acknowledges the support of the NYU Abu Dhabi Research Enhancement Fund under grant RE124. 
TMB acknowledges financial contribution from the agreement ASI-INAF n.2017-14-H.O.
AJT is supported by a Natural Sciences and Engineering Research Council of Canada (NSERC) Post-Graduate Doctoral Scholarship (PGSD2-490318-2016).
TDR acknowledges support from the Netherlands Organisation for Scientific Research (NWO) Veni Fellowship, grant number 639.041.646.
JCAM-J is the recipient of an Australian Research Council Future Fellowship (FT140101082).
The research reported in this publication was supported by Mohammed Bin Rashid Space Centre (MBRSC), Dubai, UAE, under Grant ID number 201701.SS.NYUAD. The Faulkes Telescope Project is an education partner of Las Cumbres Observatory. The Faulkes Telescopes are maintained and operated by LCO. This research has made use of the VizieR catalogue access tool, CDS, Strasbourg, France. The original description of the VizieR service was published in \cite{VizieR00}.
FCZ is supported by grant AYA2015-71042-P.
SC acknowledges  financial support from the
UnivEarthS Labex program of Sorbonne Paris Cite (ANR-
10-LABX-0023 and ANR-11-IDEX-0005-02).
 JM acknowledges support from PNHE, the OCEVU Labex (ANR-11-LABX-0060) and the A*MIDEX project (ANR-11-IDEX-0001-02) funded by the "Investissements dÕAvenir" French government program managed by the ANR. 
GRS and AJT acknowledge funding from NSERC Discovery Grants.
This work profited from discussions carried out during a meeting organized at the International Space Science Institute (ISSI) Beijing by T. Belloni and D. Bhattacharya.
The SALT observations were obtained as part of the SALT Large Science Program on transients, 2016-2-LSP-001, which is in part supported by Polish participation
in SALT funded by grant no. MNiSW DIR/WK/2016/07.
Based on observations collected at the European Southern Observatory under ESO 
programme ID 099.D-0884(A).
\end{acknowledgements}



\end{document}